\documentclass[superscriptaddress,onecolumn,english,floatfix,aps,prd,amsmath,amssymb,longbibliography,11pt]{revtex4-2}
\usepackage[T1]{fontenc}
\usepackage[latin9]{inputenc}
\usepackage{color}
\usepackage{times}
\usepackage{babel}
\usepackage{epsfig}
\usepackage{subfigure}
\usepackage{gensymb}
\usepackage{textcomp}
\usepackage{amsmath}
\usepackage{amssymb}
\usepackage{amsbsy}
\usepackage{graphicx}
\usepackage{float}
\usepackage{esint}
\usepackage[colorlinks=true]{hyperref}
\hypersetup{citecolor=blue,linkcolor=blue,urlcolor=blue}

\begin{document}

\title{Polarization and energy ellipsoids for an introductory visualization of tensors}

\author{Danilo T. Alves}
\email{danilo@ufpa.br}
\affiliation{Faculdade de F\'{i}sica, Universidade Federal do Par\'{a}, 66075-110, Bel\'{e}m, Brazil}

\author{Lucas Queiroz}
\email{lucas.queiroz@ifpa.edu.br}
\affiliation{Instituto Federal de Educa\c{c}\~{a}o, Ci\^{e}ncia e Tecnologia do Par\'{a}, Campus Marab\'{a} Rural, 68508-970, Marab\'{a}, Brazil}
\affiliation{Faculdade de F\'{i}sica, Universidade Federal do Par\'{a}, 66075-110, Bel\'{e}m, Brazil}

\author{Jeferson Danilo L. Silva}
\email{jdanilo@ufpa.br}
\affiliation{Faculdade de Engenharia da Computa\c{c}\~{a}o e Telecomunica\c{c}\~{o}es, Instituto de Tecnologia, Universidade Federal do Par\'{a}, 66075-110, Bel\'{e}m, Brazil}

\date{\today}

\begin{abstract}
In \textit{The Feynman Lectures on Physics} is discussed an introduction to tensors by means of the polarization tensor,
including a way of ``visualizing'' this tensor via the energy ellipsoid, which is drawn by the electric fields which produce the same polarization energy density in an anisotropic crystal.
Here, we discuss an alternative way of visualizing the polarization tensor, by means of the polarization ellipsoid, which is based on the ideas of Lamé's stress ellipsoid and is drawn by the 
polarization vectors produced by electric fields having the same magnitude. 
We compare both ellipsoids as a first introductory way of visualizing the polarization tensor.
\end{abstract}
%
\maketitle
%
\section{Introduction}
\label{intro}

During the early development of the theory of elasticity (1820-1830),
Cauchy and Lamé observed that the stress tensor could be visually represented by means of an ellipsoid,
named the stress ellipsoid (also known as the Lamé ellipsoid) \cite{Fung-1965}.
Nowadays, the visual representation of tensors is important in several areas,
such as in medical imaging (diffusion tensors), structural and mechanical engineering (stress and strain tensors), geophysics and seismology (stress tensors) \cite{Weickert-et-al-Book-2005,Bi-et-al-JVis-2019}.

Feynman, in Ref. \cite{Feynman-Lectures-vol-2} (Chap. 31), explains that \textit{``The mathematics of tensors is particularly useful for describing properties of substances which vary in direction -- although that's only one example of their use.''}
Then, focusing on this particular example, Feynman presents a didactic approach to introduce the discussion of tensors using the dependence of the induced polarization of an anisotropic crystal on the direction of the applied electric field:
\textit{``... one of the important properties of crystals -- or of most substances -- is that their electric polarizability is different in different directions. If you apply a field in any direction, the atomic charges shift a little and produce a dipole moment, but the magnitude of the moment depends very much on the direction of the field.''} \cite{Feynman-Lectures-vol-2}.
The relation between the applied electric field and the induced polarization is given by the polarizability tensor \cite{Feynman-Lectures-vol-2}.
Feynman also presents the energy ellipsoid as a way of ``visualizing'' the polarizability tensor. 
The energy ellipsoid is the locus of the endpoints of the electric field vectors $\bf{E}$ (originating from a common
center) which produce the same polarization energy density in an anisotropic crystal \cite{Feynman-Lectures-vol-2}.

Here, we discuss an alternative way of ``visualizing'' the polarization tensor, by means of the polarization ellipsoid, which is the locus of the endpoints of the polarization vectors $\bf{P}$ (induced dipole moment per unit volume) induced by electric fields $\bf{E}$ having a same magnitude $E$ and applied in different directions on an anisotropic crystal.
This polarization ellipsoid is based on the ideas of the Lamé's stress ellipsoid \cite{Fung-1965}.
We compare both ellipsoids, indicating the more simple for a first introductory way of visualizing the polarization tensor.

\section{The Energy Ellipsoid}
\label{sec:energy-ellip}

Let us start considering that an electric field
$\mathbf{E}$ acting on a crystal produces a polarization $\mathbf{P}$ approximately given by
\begin{equation}
	P_i=\sum_j \alpha_{ij}E_j,
	\label{eq:P-alpha-E}
\end{equation}
where: $i$ and $j$ represent either $x$, $y$, or $z$; $P_i$ and $E_j$ are the cartesian components of 
$\mathbf{P}$ and $\mathbf{E}$; and $\alpha_{ij}$ represents the elements of the tensor of polarizability \cite{Feynman-Lectures-vol-2}.
Eq. \eqref{eq:P-alpha-E} \textit{``... is a good approximation for many substances if $\mathbf{E}$ is not too large''} \cite{Feynman-Lectures-vol-2}.
In order to introduce the energy ellipsoid, Feynman starts by writing
the energy per unit volume $u_P$ required to polarize a crystal, namely,
\begin{equation}
u_P =\frac{1}{2} \mathbf{E}\cdot \mathbf{P}.
\label{eq:u_p}
\end{equation}
Thus, $u_P$ can be rewritten in terms only of the components of $\mathbf{E}$, namely,
\begin{equation}
u_P = \frac{1}{2}\sum_{i}\sum_{j}\alpha_{ij}E_{i}E_{j}.
\label{u_P}
\end{equation}

Let us consider the particular case of an electric field applied with only $x$ and $y$ components and, since $\alpha_{yx} = \alpha_{xy}$ \cite{Feynman-Lectures-vol-2}, one can obtain all values of $E_x$ and $E_y$ that correspond to a given amount of energy $u_P=u_0$ by solving the equation \cite{Feynman-Lectures-vol-2}
\begin{equation}
\alpha_{xx}E_{x}^{2}+2\alpha_{xy}E_{x}E_{y}+\alpha_{yy}E_{y}^{2}=2u_{0}.
\label{eq:ellipse}
\end{equation}
According to Feynman, this quadratic equation \textit{``... must be an ellipse, rather than a parabola or a hyperbola, because the energy for any field is always positive and finite.''} Each point of this ellipse --- the energy ellipse --- corresponds to the endpoint of an electric field $\mathbf{E}$ that polarizes the crystal with the same energy density $u_0$.
Feynman also says that \textit{``... such an ``energy ellipse'' is a nice way of ``visualizing'' the polarization tensor.''}
If all three components are included in Eq. \eqref{u_P}, the endpoint of an electric field $\mathbf{E}$ is on the surface of an ellipsoid, the so-called energy ellipsoid.

\begin{figure}
	\centering
	\subfigure[\label{fig:energy-ellipsoid-a}]{\epsfig{file=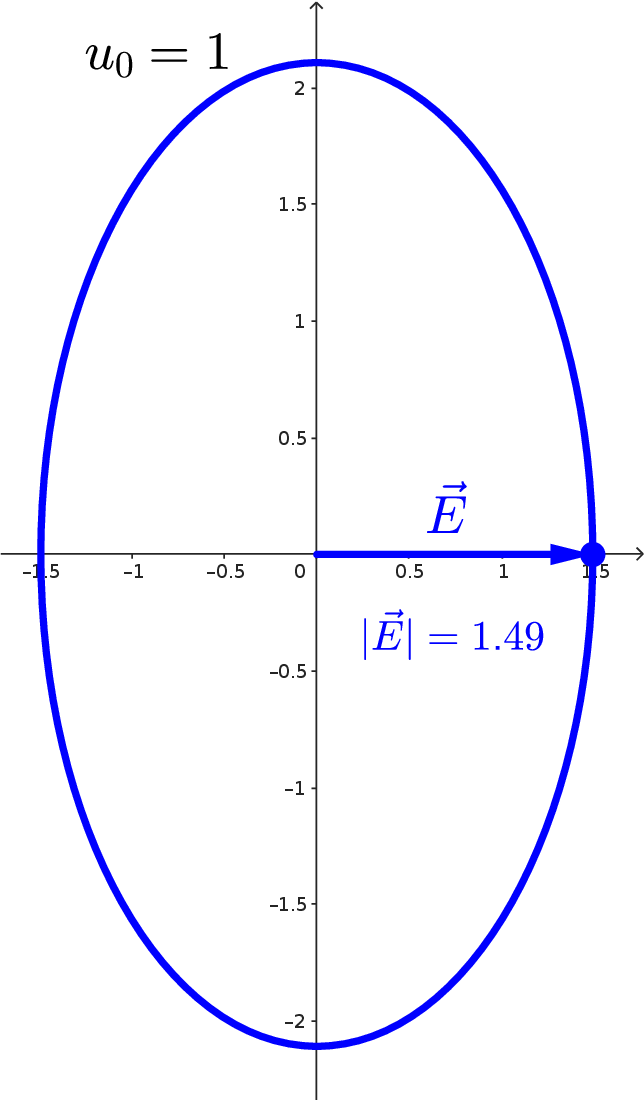, width=0.245 \linewidth}}
	\subfigure[\label{fig:energy-ellipsoid-b}]{\epsfig{file=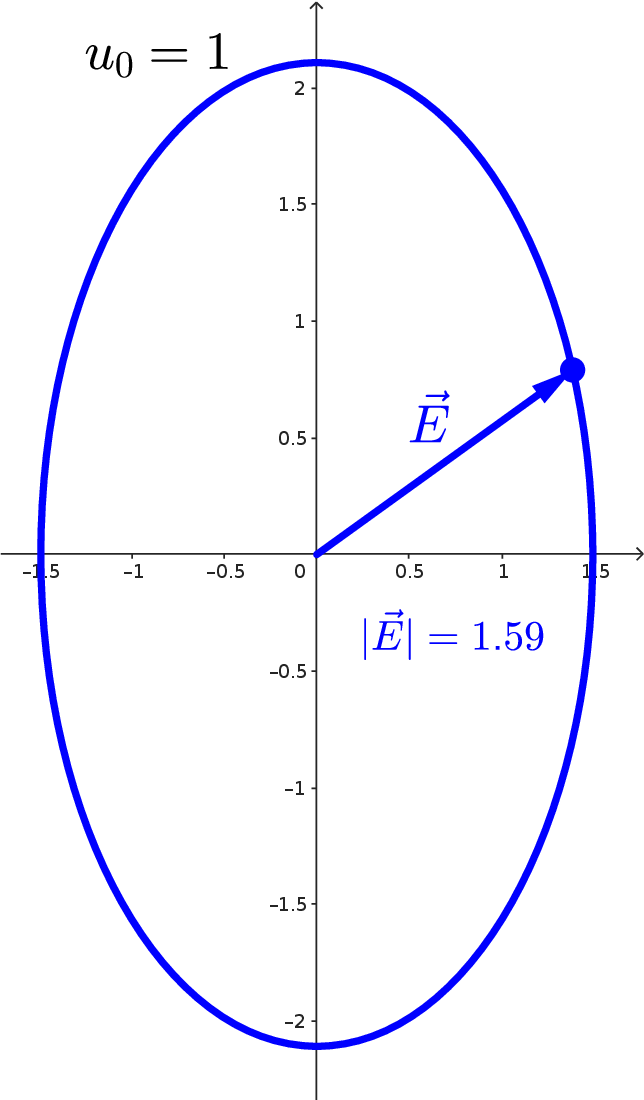, width=0.245 \linewidth}}
	\subfigure[\label{fig:energy-ellipsoid-c}]{\epsfig{file=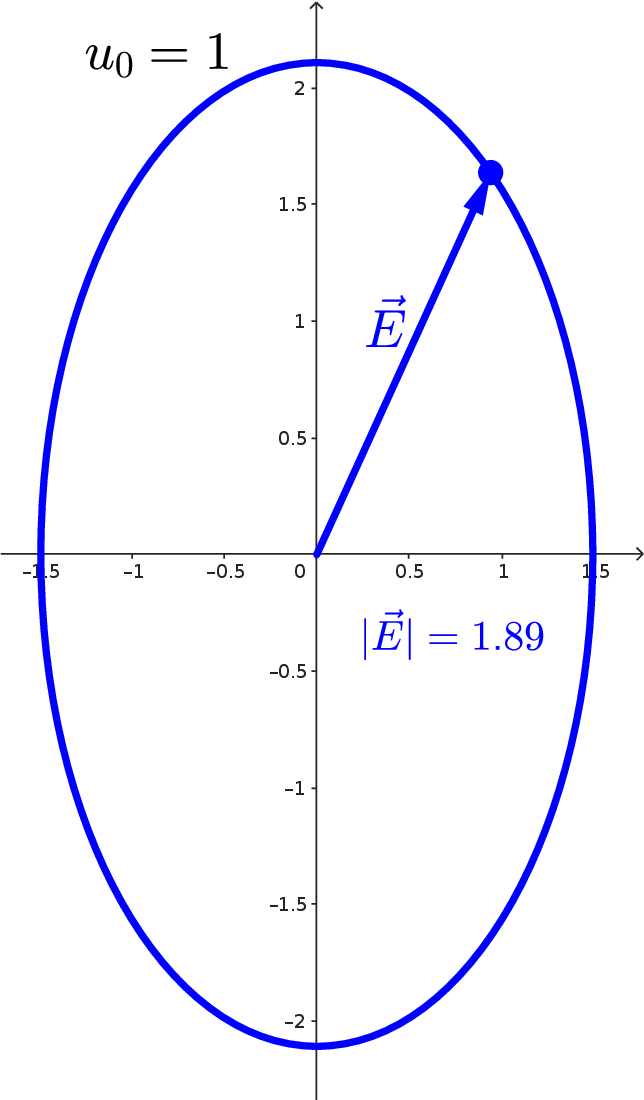, width=0.245 \linewidth}}
	\subfigure[\label{fig:energy-ellipsoid-d}]{\epsfig{file=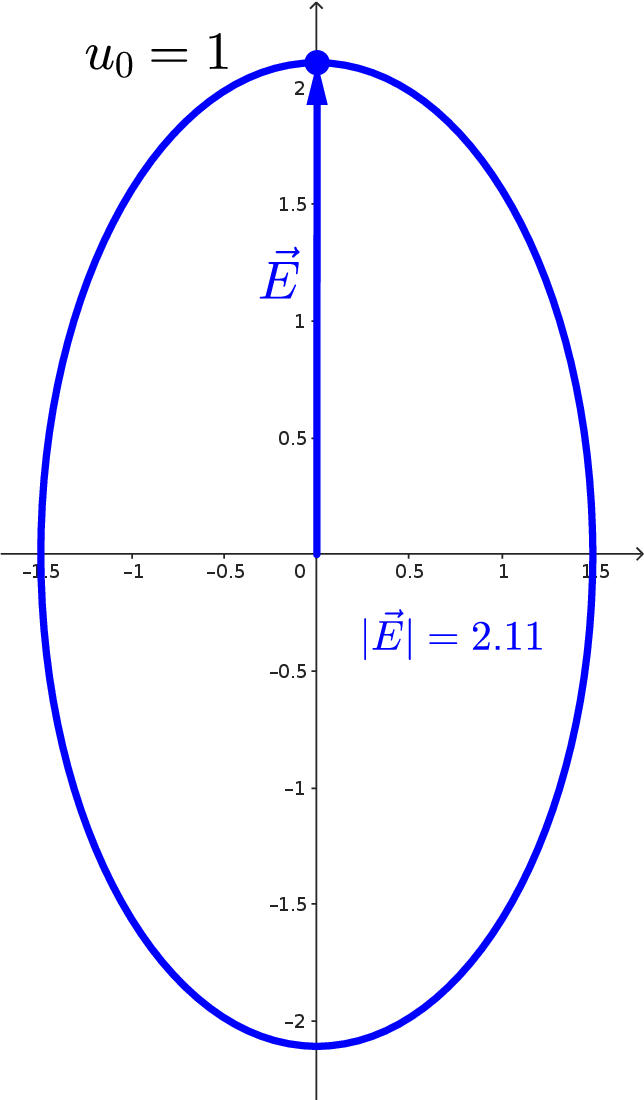, width=0.245 \linewidth}}
	\caption{
		Plots of Eq. (\ref{eq:ellipse}), considering $u_0 = 1$, $\alpha_{xx}=0.90$ and $\alpha_{yy}=0.45$ (so that $\alpha_{xx}=2\alpha_{yy}$). 
		In each plot, we show simultaneously the energy $u_0$ (the value is the same) and a vector $\mathbf{E}$ with its magnitude indicated.
		The electric fields $\bf{E}$ have different magnitudes, but they all polarize the crystal with the same energy $u_0$, and each one defines a point on the energy ellipse. 
	}
	\label{fig:energy-ellipsoid}
\end{figure}

For simplicity, let us consider $\alpha_{xy}=0$, so that Eq. \eqref{eq:ellipse} becomes
\begin{equation}
	\alpha_{xx}E_{x}^{2}+\alpha_{yy}E_{y}^{2}=2u_{0}.
	\label{eq:ellipse-2}
\end{equation}
A visual representation of this energy ellipse is given in Figs. \ref{fig:energy-ellipsoid-a}-\ref{fig:energy-ellipsoid-d},
where we show some different vectors $\mathbf{E}$, which define points on the energy ellipse, 
also indicating that the value of $u_0$ is the same in each case. 
Notice that, in the example discussed in Fig. \ref{fig:energy-ellipsoid}, the polarizability is greater in the $x$-direction ($\alpha_{xx}>\alpha_{yy}$), but the major axis of the energy ellipse is perpendicular to this direction. 
Since the energy ellipse is built by electric fields producing the same energy of polarization, its major axis coincides with the direction where the polarizability of the crystal is the smallest.
\begin{figure}[H]
	\centering
	\subfigure[\label{fig:polarization-ellipsoid-a}]{\epsfig{file=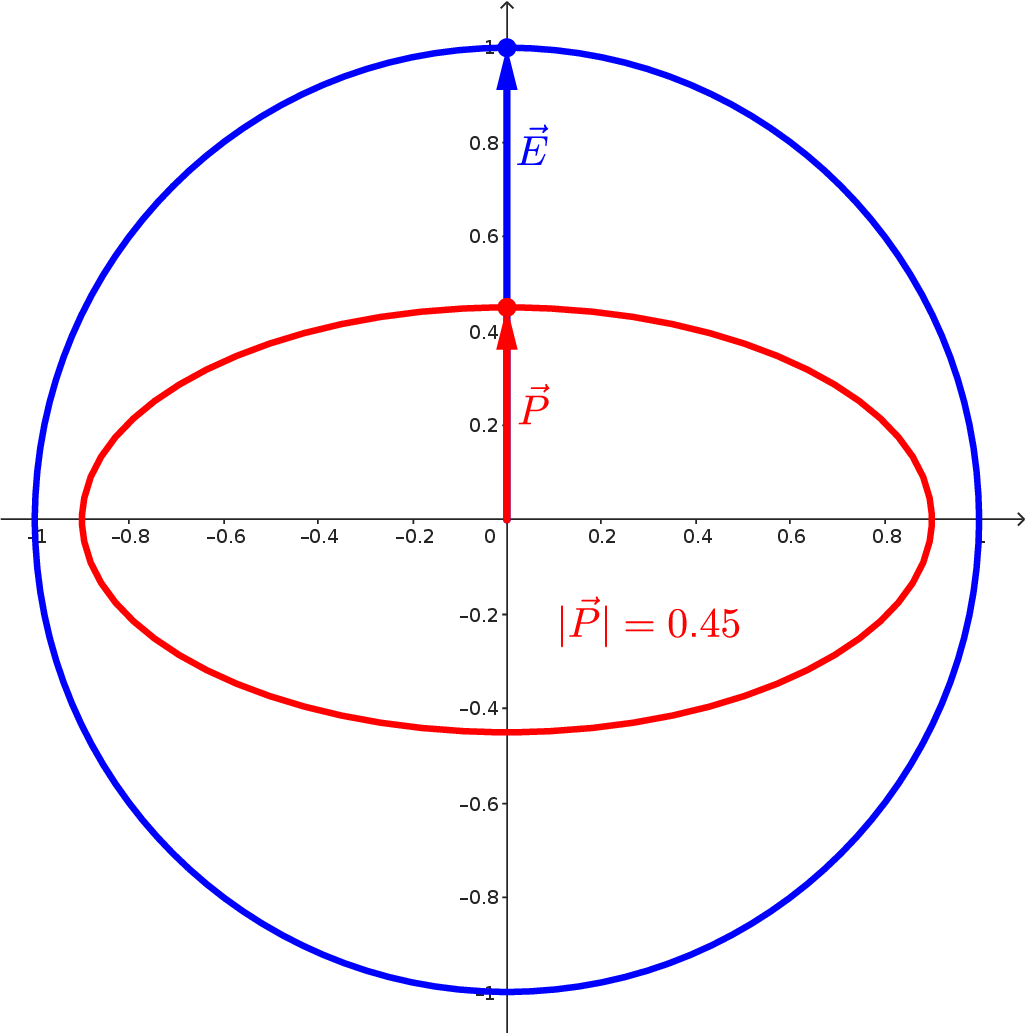, width=0.4 \linewidth}}
	\subfigure[\label{fig:polarization-ellipsoid-b}]{\epsfig{file=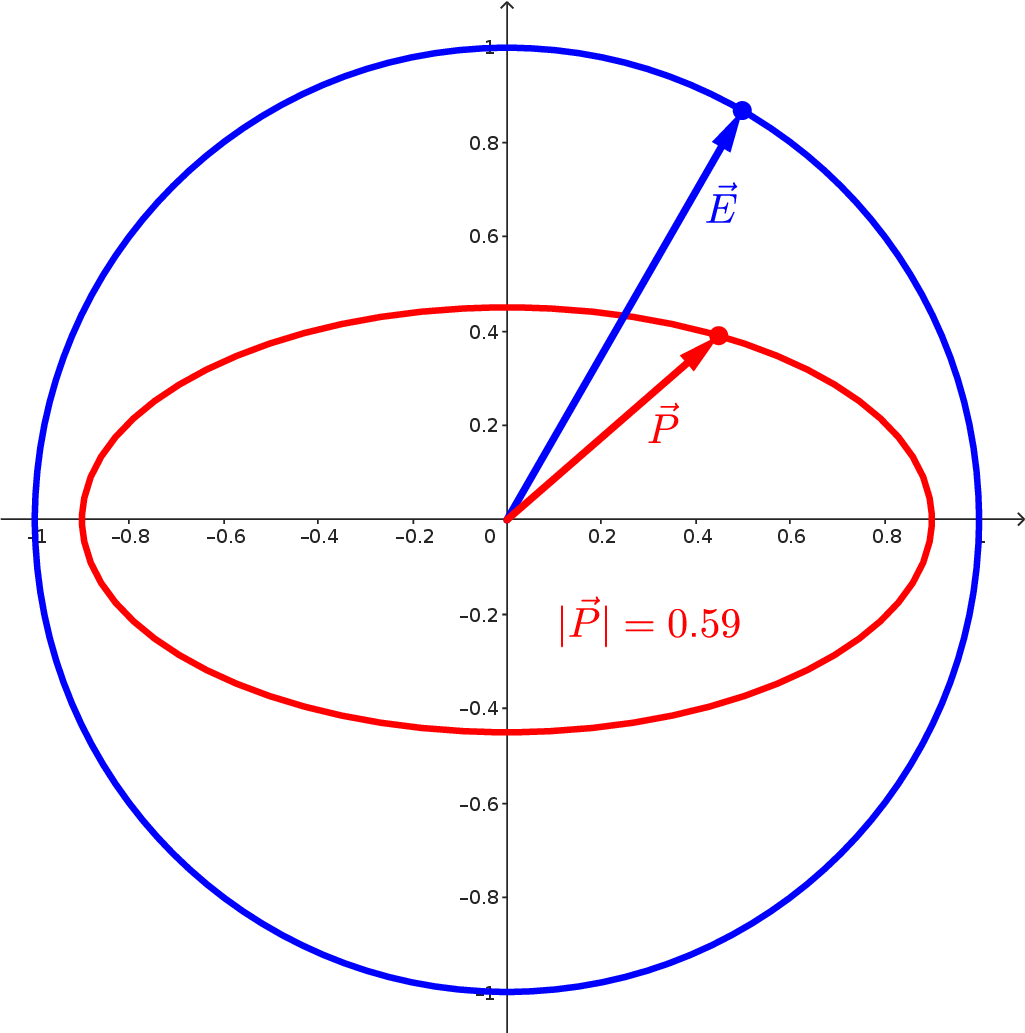, width=0.4 \linewidth}}
	\subfigure[\label{fig:polarization-ellipsoid-c}]{\epsfig{file=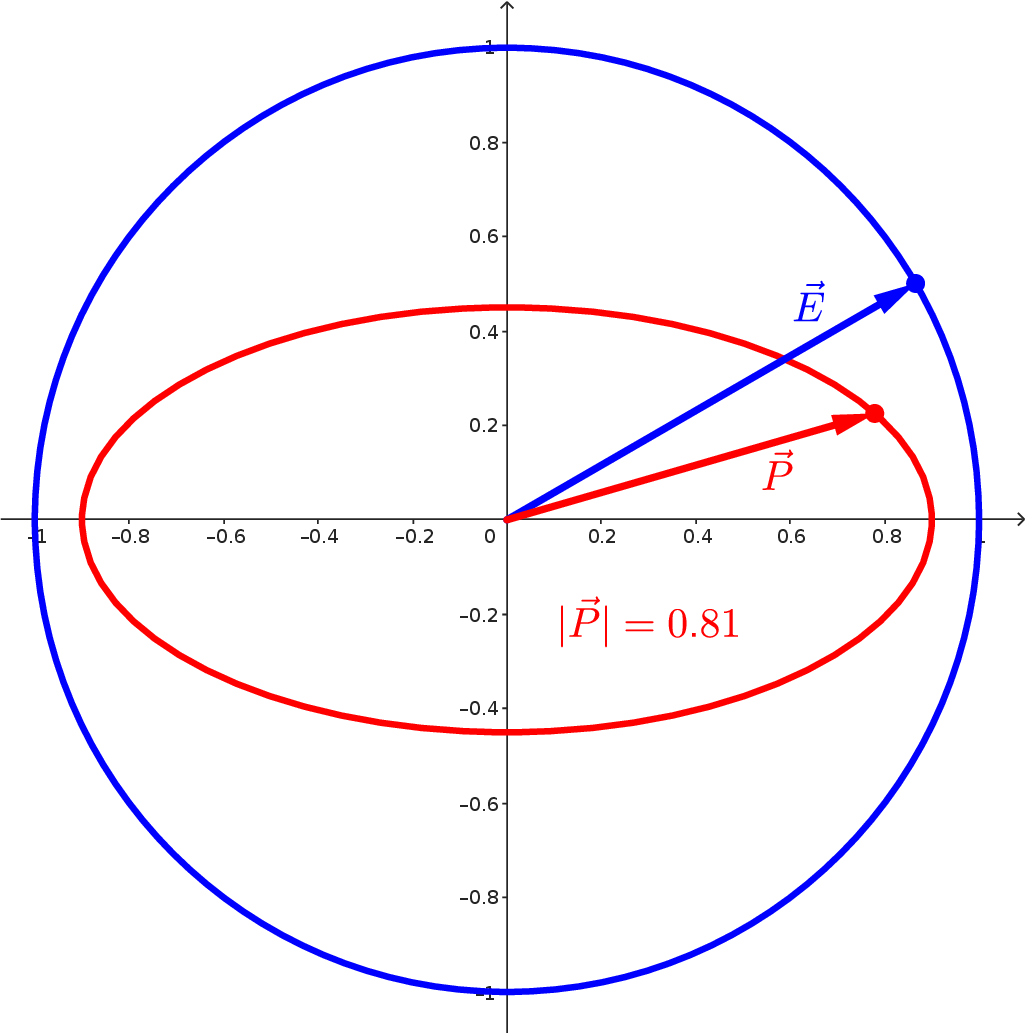, width=0.4 \linewidth}}
	\subfigure[\label{fig:polarization-ellipsoid-d}]{\epsfig{file=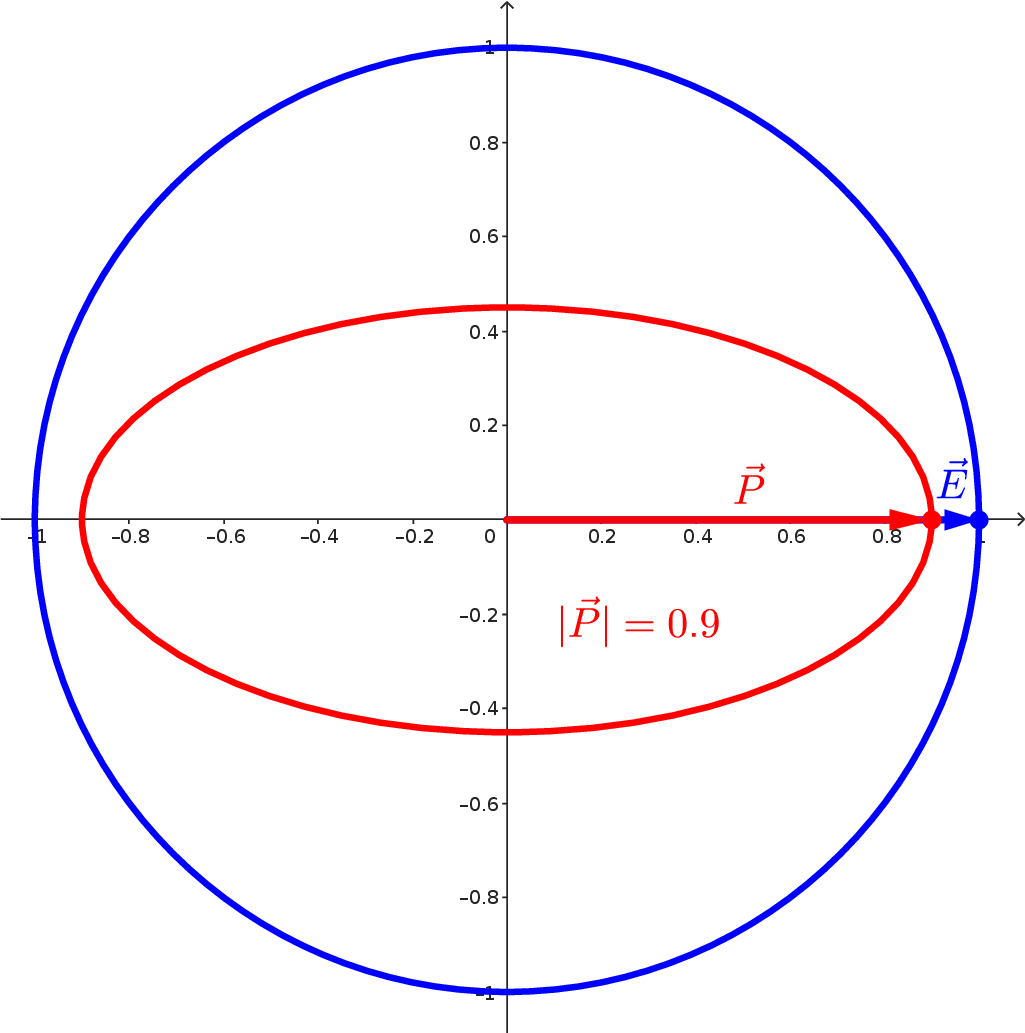, width=0.4 \linewidth}}
	\caption{
		In red lines, we show the ellipse (polarization ellipse) corresponding to Eq. (\ref{eq:ellipse-polarization-2}), considering $E=1$, $\alpha_{xx}=0.90$ and $\alpha_{yy}=0.45$ (so that $\alpha_{xx}=2\alpha_{yy})$, also showing the vectors $\mathbf{P}$ for four distinct points of this ellipse. Since in each figure $E$ is the same, the endpoints of the vectors $\mathbf{E}$ describe a circle (blue lines). In each plot, we show simultaneously a possible vector $\mathbf{E}$ and the induced polarization vector $\mathbf{P}$, with the endpoint of this latter defining a point on the polarization ellipse (red line). 
	} 
	\label{fig:polarization-ellipsoid}
\end{figure}

\section{The Polarization Ellipsoid}
\label{sec:polarization-ellip}

Now, we discuss an alternative way of ``visualizing'' the polarization tensor, 
taking as basis the ideas of the Lamé's stress ellipsoid (see Sec. 3.8 of Ref. \cite{Fung-1965}).
The main idea is to plot the vectors $\bf{P}$ via Eq. \eqref{eq:P-alpha-E}, 
but considering $E_{x}$, $E_{y}$ and $E_{z}$ in such a way that $E$ has the same magnitude.
This results in parametric equations of an ellipsoid.
Considering again the particular case of an electric field applied with only $x$ and $y$ components, we have $E^{2}=E_{x}^{2}+E_{y}^{2}$, which enable us to write $E_x=E\cos\left(\theta\right)$
and $E_y=E\sin\left(\theta\right)$. In this way, we have parametric equations of an ellipse:
\begin{eqnarray}
	P_{x}&=&\alpha_{xx}E\cos\left(\theta\right)+\alpha_{xy}E\sin\left(\theta\right),
	\label{eq:px}\\
	P_{y}&=&\alpha_{xy}E\cos\left(\theta\right)+\alpha_{yy}E\sin\left(\theta\right),
	\label{eq:py}
\end{eqnarray}
where $0\leq \theta<2\pi$. A visualization of the ellipse defined by the endpoints of $\bf{P}$,
for the case $\alpha_{xy}=0$, is given in Fig. \ref{fig:polarization-ellipsoid}, where
we show some different vectors $\mathbf{P}$ and simultaneously indicate $\mathbf{E}$ (since $E$ is the same,
the endpoints of the vectors $\mathbf{E}$ describe a circle). 
Moreover, in the example discussed in Fig. \ref{fig:polarization-ellipsoid}, the direction where the polarizability is greater ($x$-direction) coincides with the major axis of the ellipse, whereas the smaller polarization ($y$-direction) coincides with
the minor axis. 

From $E^{2}=E_{x}^{2}+E_{y}^{2}$, and using Eqs. \eqref{eq:px} and \eqref{eq:py}, we obtain 
the standard equation of the polarization ellipse:
\begin{equation}
E^{2}=\frac{\alpha_{xy}^{2}+\alpha_{yy}^{2}}{(\alpha_{xx}\alpha_{yy}-\alpha_{xy}^{2})^{2}}P_{x}^{2}-\frac{2\alpha_{xy}(\alpha_{xx}+\alpha_{yy})}{(\alpha_{xx}\alpha_{yy}-\alpha_{xy}^{2})^{2}}P_{x}P_{y}
+\frac{\alpha_{xy}^{2}+\alpha_{xx}^{2}}{(\alpha_{xx}\alpha_{yy}-\alpha_{xy}^{2})^{2}}P_{y}^{2}.	\label{eq:ellipse-polarization}
\end{equation}
If all three components are included, we have the equation of the polarization ellipsoid.
For simplicity, by considering again  $\alpha_{xy}=0$, Eq. \eqref{eq:ellipse-polarization} takes the form:
\begin{equation}
	E^{2}=\frac{1}{\alpha_{xx}^{2}}P_{x}^{2}+\frac{1}{\alpha_{yy}^{2}}P_{y}^{2}.
\label{eq:ellipse-polarization-2}
\end{equation}
which defines the polarization ellipse shown in Fig. \ref{fig:polarization-ellipsoid}.

\section{Conclusion}

Comparing the polarization and energy ellipsoids,
one can see that the former is built directly from Eq. \eqref{eq:P-alpha-E}
(by only considering  $E$ constant), whereas the latter requires, in addition
to this equation, the
introduction of the energy density given in Eq. \eqref{eq:u_p}.
In this way, the construction of the polarization ellipsoid is more direct for a first introductory visualization of the polarization tensor.
Moreover, the polarization ellipsoid provides a more direct visual interpretation, 
since the direction of the greater (lower) polarizability
coincides with the direction of the major (minor) axis of the polarization ellipsoid.


\begin{acknowledgments}

The authors thank Marcelo C. de Lima, Nelson P. C. de Souza, and Van Sérgio Alves for fruitful discussions.
L.Q. was supported by the Coordenação de Aperfeiçoamento de Pessoal de Nível Superior - Brasil (CAPES), Finance Code 001.
This work was partially supported by CNPq - Brazil, Processo 408735/2023-6 CNPq/MCTI.
\end{acknowledgments}
%

%

\end{document}